\documentclass[
reprint,
physrev,
]{revtex4-1}
\usepackage{amsmath} %this allows use of Equation mode
\usepackage[T1]{fontenc}
\usepackage[latin1]{inputenc}
\usepackage{amssymb}
\usepackage{setspace}
\usepackage{color}
\usepackage{float}
\usepackage[english]{babel}
\usepackage{dcolumn}% Align table columns on decimal point
\usepackage{bm}% bold math
\usepackage{txfonts}               % free Times-based fonts for text and math
\usepackage{graphicx}   

\newcommand{\mody}{\color{black}}

\newcommand{\norm}{\color{black}}

\begin{document}

\title{Sandpile modelling of pellet pacing in fusion plasmas}

\author{C. A. Bowie}
\email{craig.bowie@anu.edu.au}
\affiliation{Australian National University, Canberra, ACT 0200, Australia}

\author{M. J. Hole}
\affiliation{Australian National University, Canberra, ACT 0200, Australia}

Copyright American Physical Society

\begin{abstract}

\mody

Sandpile models have been used to provide simple phenomenological models without incorporating the detailed features of a fully featured model. The Chapman sandpile model (Chapman \textit{et al} \textit{Physical Review Letters} 86, 2814 (2001)) has been used as an analogue for the behaviour of a plasma edge, with mass loss events being used as analogues for ELMs. In this work we modify the Chapman sandpile model by providing for both increased and intermittent driving. We show that the behaviour of the sandpile, when continuously fuelled at very high driving, can be determined analytically by a simple algorithm. We observe that the size of the largest avalanches is better reduced by increasing constant driving than by the intermittent introduction of `pellets' of sand. Using the sandpile model as a reduced model of ELMing behaviour, we conject that ELM control in a fusion plasma may similarly prove more effective with increased total fuelling than with pellet addition.

\norm

\end{abstract}

\maketitle

\section{Introduction \label{sec:Introduction}}
% \vspace{-0.25cm}

Pellet injection has been extensively used as a candidate for ELM control and reduction in fusion plasmas.~\cite{Hong1999,Baylor2005,Baylor2007,Baylor2013,Baylor2015,Lang2004,Lang2014,Lang2015,Pegourie2007,Rhee2012} Pellet size, frequency, and location have all been tested experimentally on ASDEX Upgrade,~\cite{Lang2004,Lang2015, Lang2018} DIII-D,~\cite{Baylor2005,Baylor2013} JET,~\cite{Baylor2015, Lang2011, Lang2015} and EAST~\cite{Li2014,Yao2017} and ELM control using pellets is being considered for use in ITER.~\cite{Doyle2007,Baylor2015}

One way of addressing the impact of pellet injection on both confinement and ELM behaviour is to seek to identify a physical system whose relaxation processes have characteristics similar to those of the ELMing process under consideration. Of particular interest is the sandpile,~\cite{Bak1987} whose relevance to fusion plasmas is well known.~\cite{Chapman1999,Dendy1997}

\begin{figure}
\centering
\includegraphics[width=1\linewidth]{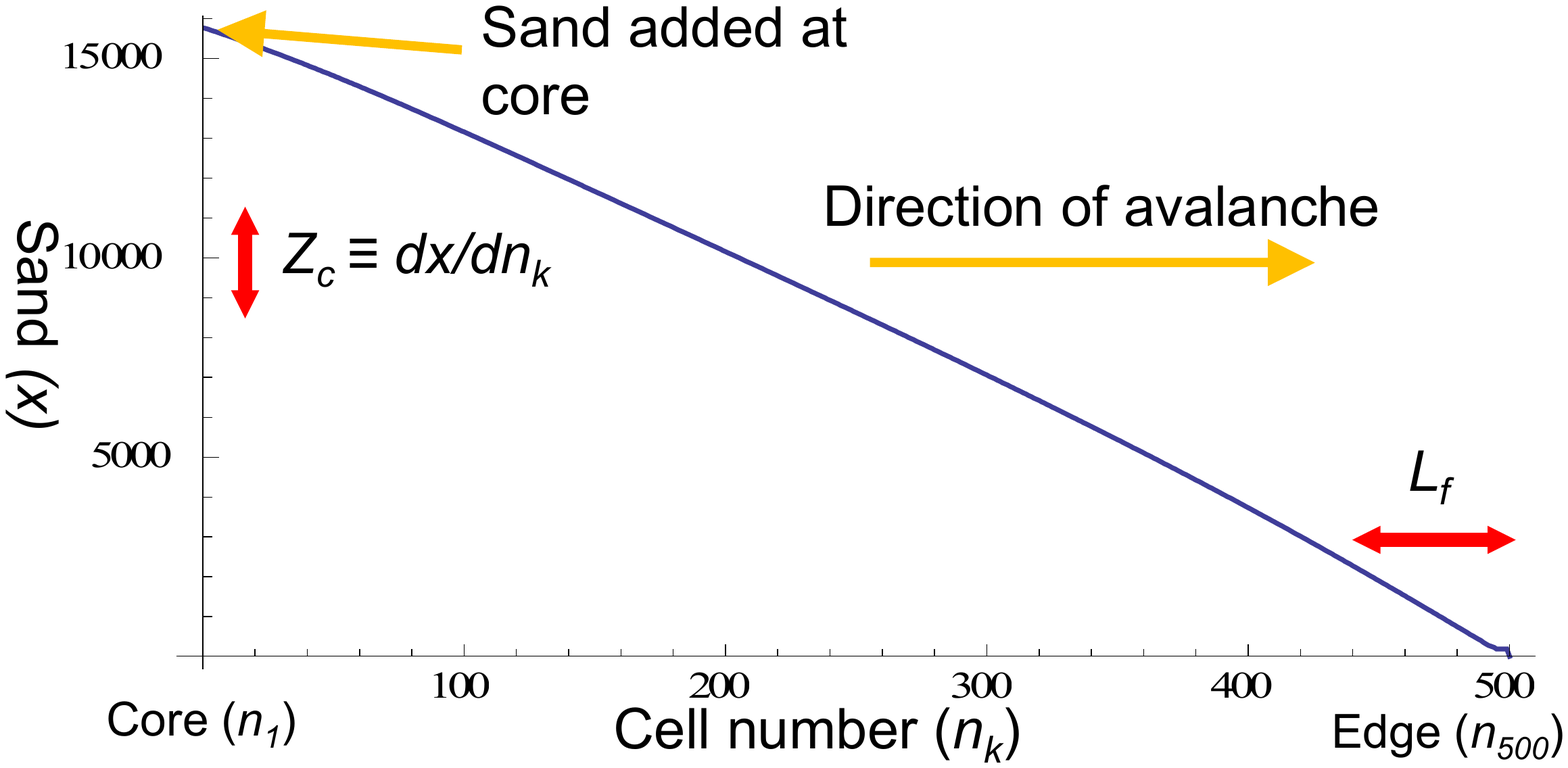}
\caption[Sandpile schematic]{Sandpile schematic, showing the key features of the sandpile model discussed in this paper. The schematic is the same for both the classic and running models, which differ only in respect of whether further sand is added only after an avalanche has concluded (classic model) or during an avalanche (running model).}
\label{fig:sandpile-schematic}
\end{figure}

Sandpile models generate avalanches, which may either be internal or systemwide, in which case particles are lost from the system. These avalanches are the response to steady fuelling of a system which relaxes through coupled near-neighbour transport events that occur whenever a critical gradient is locally exceeded. The possibility that, in some circumstances, ELMing may resemble avalanching was raised~\cite{Chapman2001A} in studies of the specific sandpile model of Ref.~\cite{Chapman2000}, a schematic of which is given in Figure \ref{fig:sandpile-schematic}. This simple one-dimensional N-cell sandpile model~\cite{Chapman2000,Chapman2001A} incorporates other established models~\cite{Bak1987,Dendy1998A} as limiting cases. It is centrally fuelled at cell \mody$n = 1$\norm, and its distinctive feature is the rule for local redistribution of sand near a cell (say at $n = k$) at which the critical gradient $Z_{c}$ is exceeded. The sandpile is conservatively flattened around the unstable cell over a fast redistribution lengthscale $L_{f}$, which spans the cells $n = k - (L_{f}  - 1), k - (L_{f} - 2), ... , k+1$, so that the total amount of sand in the fluidization region before and after the flattening is unchanged. Because the value at cell $n = k+1$ prior to the redistribution is lower than the value of the cells behind it, the redistribution results in the relocation of sand from the fluidization region, to the cell at $n = k + 1$. If redistributions are sequentially triggered outwards across neighbouring cells, leading to sand ultimately being output at the edge of the sandpile, an avalanche is said to have occurred. The sandpile is then fuelled again, either after the sandpile has iterated to stability so that sand ceases to escape from the system (`classic model'), or immediately after the first `sweep' through the system has been completed (`running model').

The lengthscale $L_{f}$, normalized to the system scale $N$, is typically ~\cite{Chapman1999,Chapman2001A,Chapman2001B,Chapman2003,Chapman2004} treated as the model's primary control parameter $L_{f}/N$, which governs different regimes of avalanche statistics and system dynamics. The lengthscale is constant across the sandpile in the classic and running models. Table \ref{Table:key parameters} summarises the key features of the model, along with the maximum and minimum values used for each variable in this paper. As will be observed from Table \ref{Table:key parameters}, we are primarily concerned here with variation in driving.
\mody
\begin{table}
\begin{tabular}{|l|c|r|}
\hline 
Parameter & Meaning & Range \\ 
\hline 
$Z_c$ & Critical gradient &100-120 \\ 
\hline 
$L_f$ & Fluidization length &5-6 \\ 
\hline 
$dx$ & Drive &1.2-4100\\ 
\hline 
$N$ & Total number of cells&500 \\\hline
\end{tabular} 
\label{Table:key parameters}
\caption{Key parameters}
\end{table}
\norm
Unlike some,~\cite{Chapman1999,Chapman2001A,Chapman2001B,Chapman2003,Chapman2004} but not all,~\cite{Bowie2016,Bowie2018} implementations of the Chapman model, $Z_{c}$ is single valued, rather than being randomized. The phenomenology generated by this model has several features resembling tokamak plasmas, including edge pedestals, enhanced confinement~\cite{Chapman2001A} and self-generated internal transport barriers~\cite{Chapman2003}. Particularly relevant here are the systemwide avalanches, or MLEs, resulting (unlike the more numerous internal avalanches which are not considered here) in mass loss from the sandpile. \mody In particular, we have focused on the MAX MLE size, being the amount of mass lost in the largest avalanches.\norm

In the `classic' sandpile model, the avalanche may propagate through the sandpile multiple times until the system ceases to output sand, prior to further fuelling of the sandpile. Effectively, fuelling is paused until the system is stable, which reflects the instantaneous nature of an avalanche, by comparison to the slow addition of single grains of sand. In the `running model' which was first explored by Bowie, Dendy and Hole~\cite{Bowie2016}, the sandpile is fuelled again as soon as the first iteration of the avalanche is complete, while the sandpile remains in a critical state. In the low fuelling regime, little difference is observed between the classic and running models: the sandpile may take $\approx500$ iterations to reach stability, during which time enough sand has been added at the first cell to cause the critical gradient to be exceeded between the first and second cells a further one or two times. Compared to the total amount of sand which may be lost in the continuing avalanche (up to $3\times10^5$ units of sand), the further sand added during the avalanche is of little relevance, being less than 1\% of the sand lost during the avalanche. By comparison, if a high fuelling rate is employed, then the extra sand added during the continuing avalanche becomes significant, and can significantly change the overall behaviour of the running model.

Typically, sandpile models are analysed in the low driving regime, as low driving is considered to be necessary to achieve a separation of time scales which is a condition of self-organized criticality (SOC) ~\cite{Watkins1999}. High driving has also been considered in relation to the Chapman model ~\cite{Watkins1999,Chapman2004} and been found to lead to the elimination of the smallest scale avalanches. Further, an analogy between the Reynolds number, and the relationship between driving and dissipation has been identified  and found to give a means of distinguishing between turbulence and SOC~\cite{Chapman2009A,Chapman2009B}. In this study, we have focused on the high driving regime, and its relationship to the total potential energy of the system, and to mass loss events (MLEs), rather than focusing on the relationship between driving and SOC.

\mody Here we make an assumption that the sandpile model is relevant to analysis of the ELMing behaviour of a fusion plasma. While this assumption is supported at low driving rates by the work of Chapman and Dendy~\cite{Chapman1999,Dendy1997}, we here seek to extend the analogy to high driving behaviour in the sandpile. \norm Specifically, we seek to draw comparisons with pellet pacing at the core of a fusion plasma by varying the amount of sand, $dx$, added at each \mody iteration, or \norm  time step. We do this in two separate ways: by setting a high constant $dx$ in order to move into the high fuelling regime; and by varying $dx$ intermittently (i.e. adding pellets) to seek to trigger avalanches. By doing this, we are able to compare systems where `pellets' are added at each time step before the system has an opportunity to fully relax (using high constant $dx$ in the running model), and systems in which the system can fully relax between `pellets', using the classic model at low fuelling with the intermittent addition of `pellets'. Using high constant $dx$ gives a proxy to pellet fuelling at the core if the pellet size is sufficient that it continues to be ablated during the occurrence of an avalanche.

We also briefly comment on the behaviour of the classic model at high fuelling, although our focus is on the behaviour of the running model at high fuelling. Finally, we consider the behaviour of the running model at extremely high driving, where the shape of the resultant sandpile is determined by a simple algorithm.

We observe that there is no single relationship between driving, waiting times, and potential energy, which holds in all regimes. Further, the nature of the relationship is different for the classic model and the running model. We comment here on the different relationships in different driving regimes, and offer insights on the reasons for those relationships, and postulate whether there are real world scenarios which may be informed by those reasons.

\section{Intermittent extra sand - Core fuelling\label{intermittent extra sand}}

We have taken the classic model, and added extra sand (pellets) in various combinations of intervals and pellet size, by way of comparison to pellet pacing in fusion plasmas. In each case, pellets are added at the core, consistent with the `ordinary' fuelling location. 

We employ only the classic model for this purpose - as discussed in Section \ref{sec:High driving - constant fuelling}, low to medium increases in constant driving do not affect $E_p$ in the classic model, while they do in the running model. Employing the running model would add a confounding variable, namely the model specific effect of the increase in total driving, as opposed to the effect of the addition of pellets, although we note that the model specific effect would be quite minor at the pellet sizes and times discussed here, as total fuelling is increased only by about a factor of two.

Lang \cite{Lang2015} discussed pellets added at lower frequencies (higher waiting time between pellets, $\Delta t_P$) with pellet timing aligned to ELM onset. These pellets triggered ELMs. Lang\cite{Lang2015} observes that as pellets increase the plasma density, this in turn increases the L-H threshold.

We have tested these observations against our model. We observe that while potential energy ($E_p$, given here by the sum of the squares of the cell values), used here as a proxy for plasma pressure, increases with pellet size, maximum MLE size (i.e. the number of grains of `sand' lost in the largest avalanche) also increases, and at a faster rate. 

For this purpose, there is a close relationship between the high driving regime discussed in section \ref{sec:driving}and the intermittent addition of extra sand. If the extra sand is absorbed into the sandpile without triggering an MLE, then the addition of the extra sand may serve simply to increase fuelling of the system. On the other hand, if the intermittent addition of extra sand triggers an MLE when the extra sand is added, the system may behave quite differently. 

Three waiting times are considered here - the waiting time between addition of pellets, $\Delta t_P$, the `natural' waiting time between MLEs for a given amount of constant fuelling (including pellet fuelling), $\Delta t_N$, and the actual waiting time observed (including pellet fuelling), $\Delta t_A$. For this purpose, $\Delta t_P$ and $\Delta t_A$ are determined by identifying the primary peak in the resulting probability distribution function (pdf) of waiting times between MLEs in the particular scenario. These times will be equal to each other if the pellets do nothing more than add to total fuelling. They differ if pellet fuelling triggers `shocks' to the system, triggering an immediate MLE before the system would otherwise have reached a critical state if the total fuelling had been constant.

For `macro' pellets, we show here results for two values of $\Delta t_P$ - $70000$ and $100000$.  Short and long $\Delta t_N$ are typically \cite{Bowie2016} observed in the model: the `short' $\Delta t_N$ for this model with $dx=1.2$ is $\approx 70000$, with a longer $\Delta t_N$ at $\approx 140,000$. The  values of $\Delta t_P$ selected therefore represent different stages in build-up prior to, or post, avalanching (recognizing that the additional fuel added by way of `pellets' also increases total fuelling).

We observe that the amount of sand lost during the largest MLE is roughly equal to double the amount of material added during the longest waiting time. As a result, if the longest waiting time remains approximately constant while the amount of material added per unit time increases (due to the introduction of pellets), then the maximum MLE size increases.

\begin{figure}%
\centering
\begin{minipage}{\linewidth}%
\includegraphics[clip,trim=0cm 0cm 0cm 0cm,width=\linewidth]{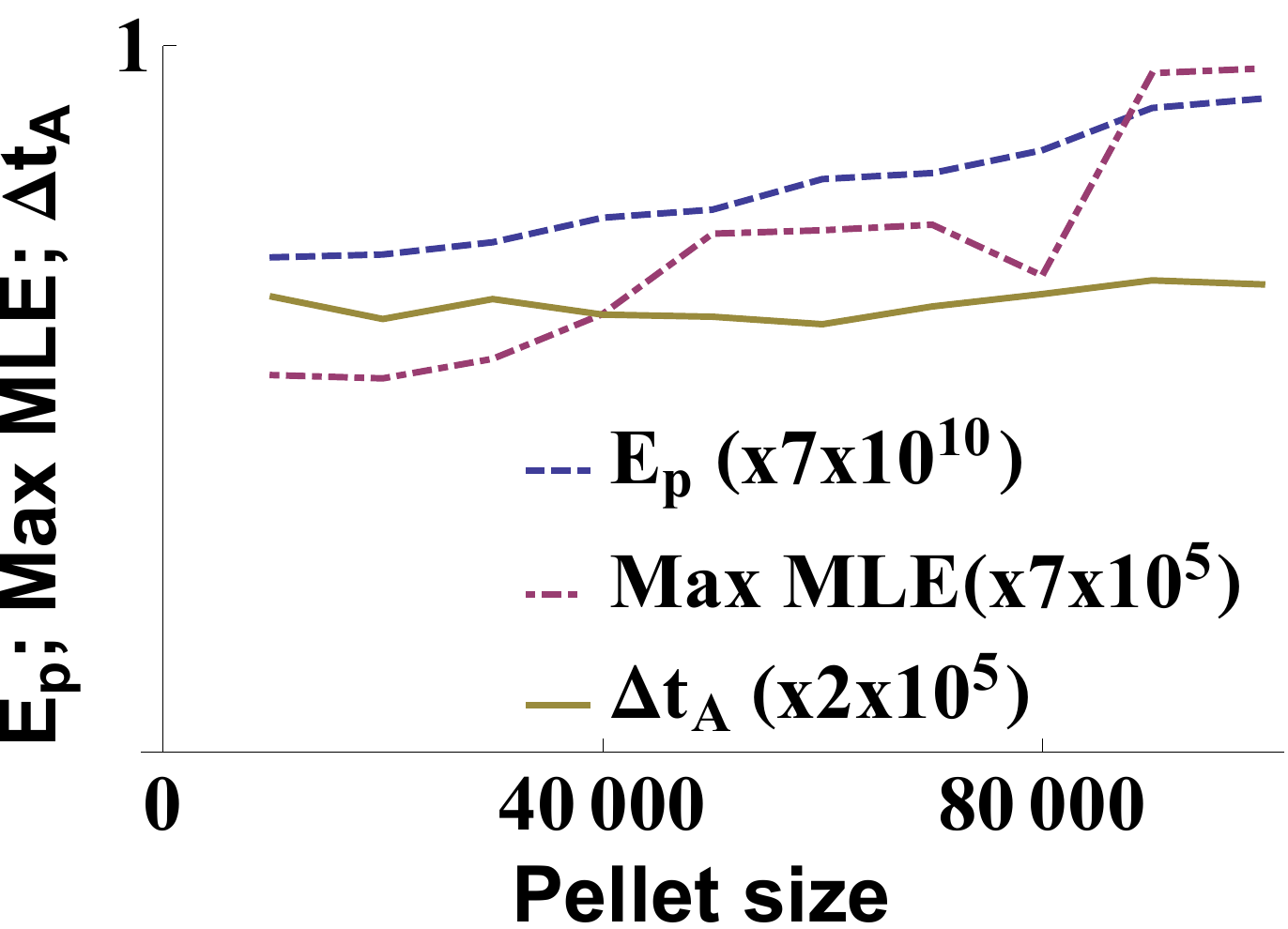}
\end{minipage}
\begin{minipage}{\linewidth}%
\includegraphics[clip,trim=0cm 0cm 0cm 0cm,width=\linewidth]{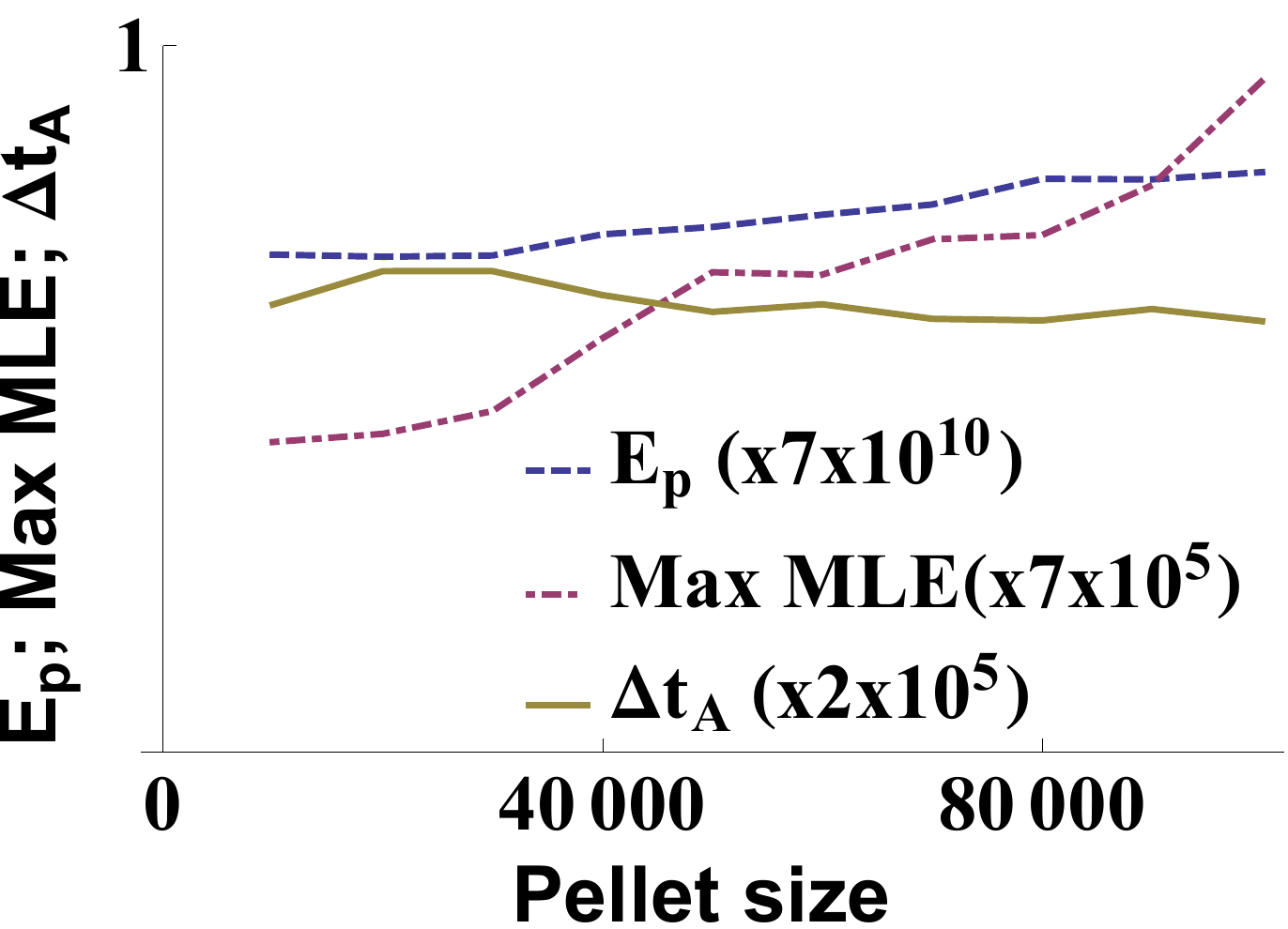}
\end{minipage}
\caption[time between pellets 70,000 and 100,000]{$E_p$, max MLE, and max $\Delta t_A$, for varying pellet sizes, with $\Delta t_P=70000$ (top); $\Delta t_P=100000$ (bottom). In all cases, fuelling occurs at the core.  }
\label{fig:time-between-pellets---70000-cell-1}
\end{figure}

Figure~\ref{fig:time-between-pellets---70000-cell-1} shows, for $\Delta t_P=70000$ and $\Delta t_P=100000$, both potential energy and maximum MLE size increase, with increasing pellet size. Maximum $\Delta t_A$ is constant in each case, although slightly longer for $\Delta t_P=100000$, which is consistent with the fact that driving is higher due to the higher pellet frequency where $\Delta t_P=70000$. In both cases, the general trend is that max MLE size increases with increasing pellet size at a faster rate than $E_p$. \mody We note that a pellet size of $80000$ represents ~2\% of the total number of grains in the sandpile.\norm

It is apparent that, at least in this model, adding pellets at the core does not, in general, reduce MLE size. In order to reach the threshold for triggering MLEs with the addition of each pellet, pellets must be so large that the resulting MLE is of a greater size than `natural' MLEs (as shown by the increasing MLE size in Figure \ref{fig:time-between-pellets---70000-cell-1}(a)and (b)), and, further, that pellets become a significant component of total fuelling. For example, for a pellet size of $70,000$, with $\Delta t_P=70000$, average total fuelling increases from $dx=1.2$ to $dx=2.2$. As a result, pellet fuelling at the core is not effective to reduce MLE size in this model. 

\section{High driving - constant fuelling \label{sec:High driving - constant fuelling}}

We now consider the impact of increasing constant fuelling, primarily in the running model. We have also briefly considered, in Section \ref{very high driving - running}, increasing constant fuelling in the classic model.

Before commenting on the high driving regimes, we first comment on some relationships observed at low and medium driving, for the purposes of observing the changes in those relationships as driving increases. For all examples, $Z_c=120$, meaning that for $dx=1.2$, $dx/Z_c=0.01$.

\begin{figure}
\centering
\includegraphics[width=1\linewidth,trim={0cm, 0cm, 0cm, 0cm}, clip]{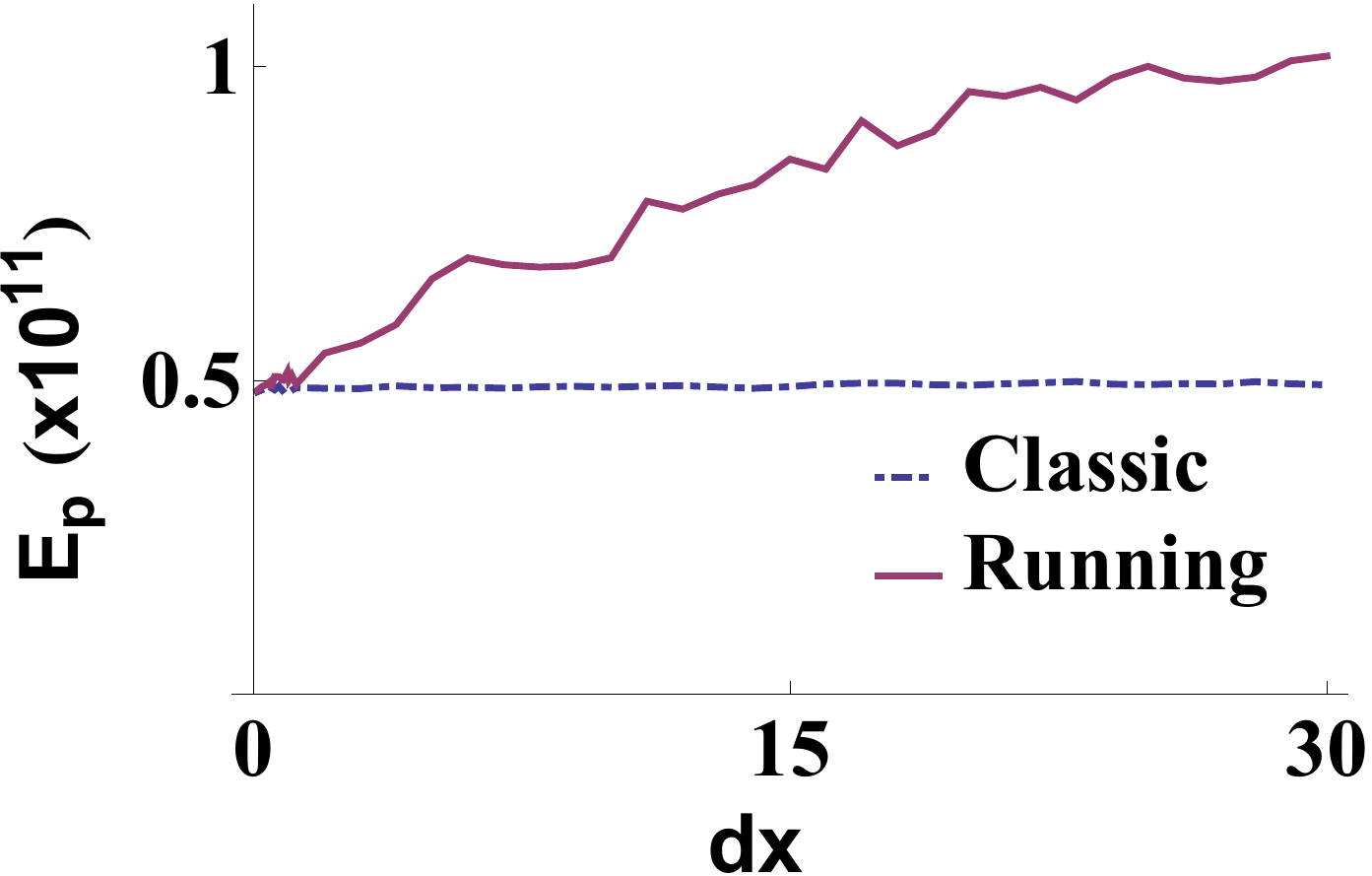}
\caption[PE versus dx - high dx]{$E_p$ versus $dx$ up to $dx = 30$, for classic and running models. It is notable that $E_p$ is effectively constant for all values of $dx$ shown here in the classic model, while $E_p$ gradually increases in the running model.}
\label{fig:Potentialenergyversusdx-highdx}
\end{figure}

We first consider changes in the driving regime for the classic and running models up to $dx = 30$. Figure \ref{fig:Potentialenergyversusdx-highdx} shows that the classic and running models produce very similar results in terms of $E_p$ at low $dx$, but vary significantly above $dx\approx 2$. We explore in subsequent sections the reasons why $E_p$ changes with $dx$ at low-medium $dx$ in the running model, but not the classic model.

\subsection{Relationship between driving and potential energy - running model \label{sec:driving}}
% \vspace{-0.25cm}

We now turn to consider the behaviour of the running model at high constant driving. Unlike the classic model, significant changes in behaviour of the system are observed as driving increases in the running model, until finally, at very high driving, the amount of sand entering and leaving the system at each iteration equalises. While we comment in section \ref{very high driving - running} below on the reasons for this behaviour at very high driving, our attention is primarily drawn to the behaviour of the system as driving increases up to that point.

\begin{figure}
\centering
\includegraphics[trim= {0mm 0mm 0mm 0mm},clip,width=\linewidth]{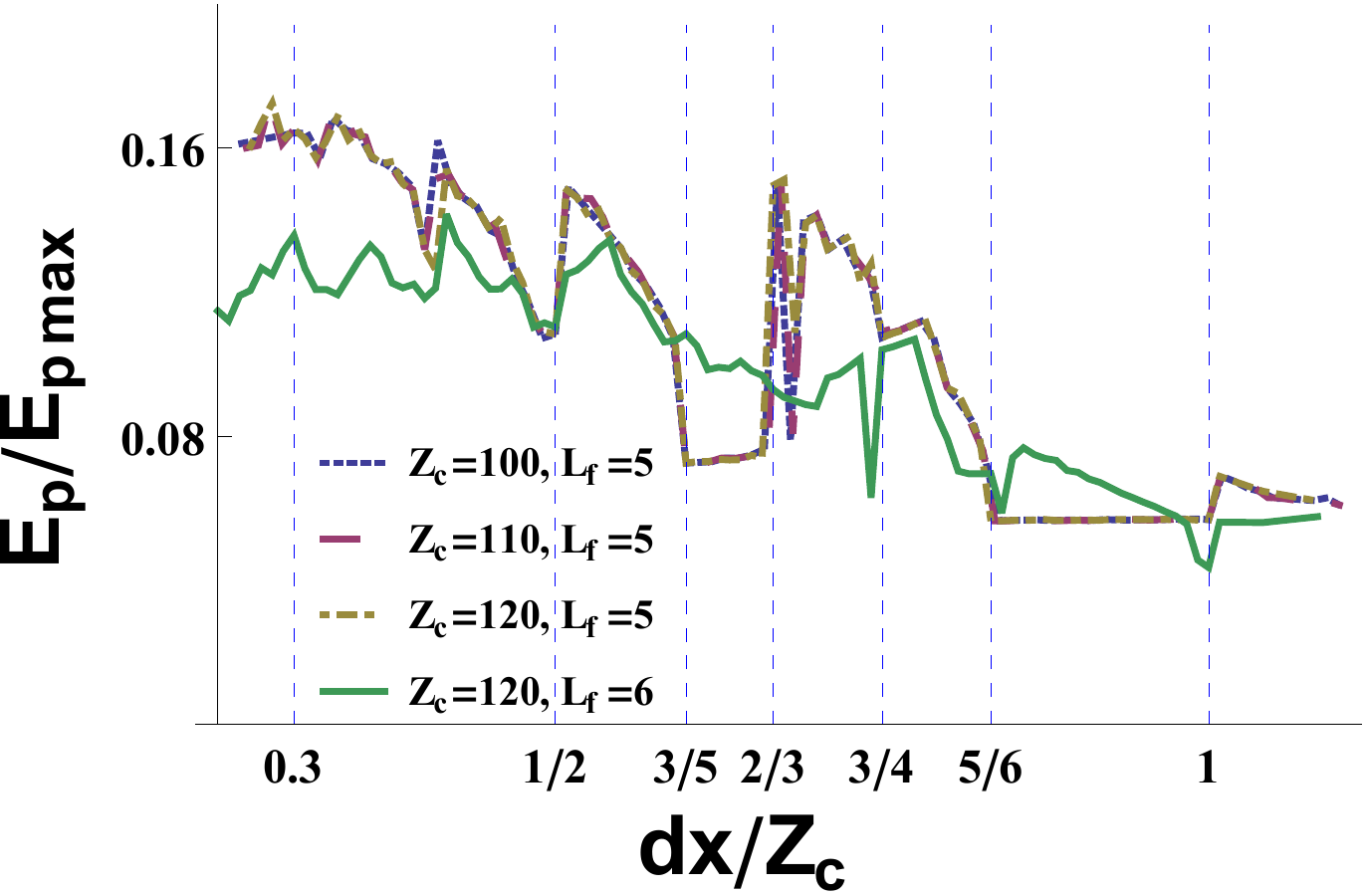}
\caption[PE/MaxPE]{Potential energy/Maximum potential energy vs $dx/Z_{c}$ (all unitless). The potential energy measured is the average potential energy (given by the sum of the squares of the cells) after the system has evolved from a nil sandpile to a `steady state', which typically takes several hundred thousand iterations. The maximum $E_p$ is calculated on the basis that actual gradient is equal to $Z_c$, i.e that the sandpile is in a maximally critical state. The three curves which largely coincide represent data for different values of $Z_{c}$, but common values of $L_{f}$. The other curve represents data for a changed value of $L_{f}$.}
\label{fig:PE_MaxPE}
\end{figure}

We show, in Figure~\ref{fig:PE_MaxPE}, $E_p/E_{p max}$ against $dx/Z_c$ for four different sets of values of $Z_c$ and $L_f$. A clear upward trend is observable for $dx/Z_c$ up to about $0.3$, with a subsequent general decline, subject to significant detailed structure. In Figure~\ref{fig:PE_MaxPE}, we focus on this region of detailed structure. It is notable that fine structure, i.e. abrupt changes in behaviour, is seen around integer ratios of $dx/Z_{c}$.

The primary peak is situated at approximately $dx/Z_c=0.3$, which is to say that the energy of the sandpile is maximised if the amount of sand added at each timestep is about $1/3$ of that sufficient to provoke an avalanche (assuming an otherwise nil gradient at the top of the sandpile). In most cases, the avalanche will not be systemwide, but will terminate before it reaches the edge - if all avalanches reached the edge, then $\Delta t_N$ would $\approxeq$ $Z_c/dx$. Ignoring fine structure, there is a systematic decrease of Max MLE with increasing $dx$ over the range $0<dx<60$. Figure \ref{fig:PE_MaxPE} demonstrates that the fine structure relates to integer ratios of $dx/Z_{c}$. Further, the shape of this potential energy curve is unchanged with variations in $Z_c$, although it is not constant with changes in $L_f$.

\begin{figure}
\centering
\includegraphics[width=\linewidth]{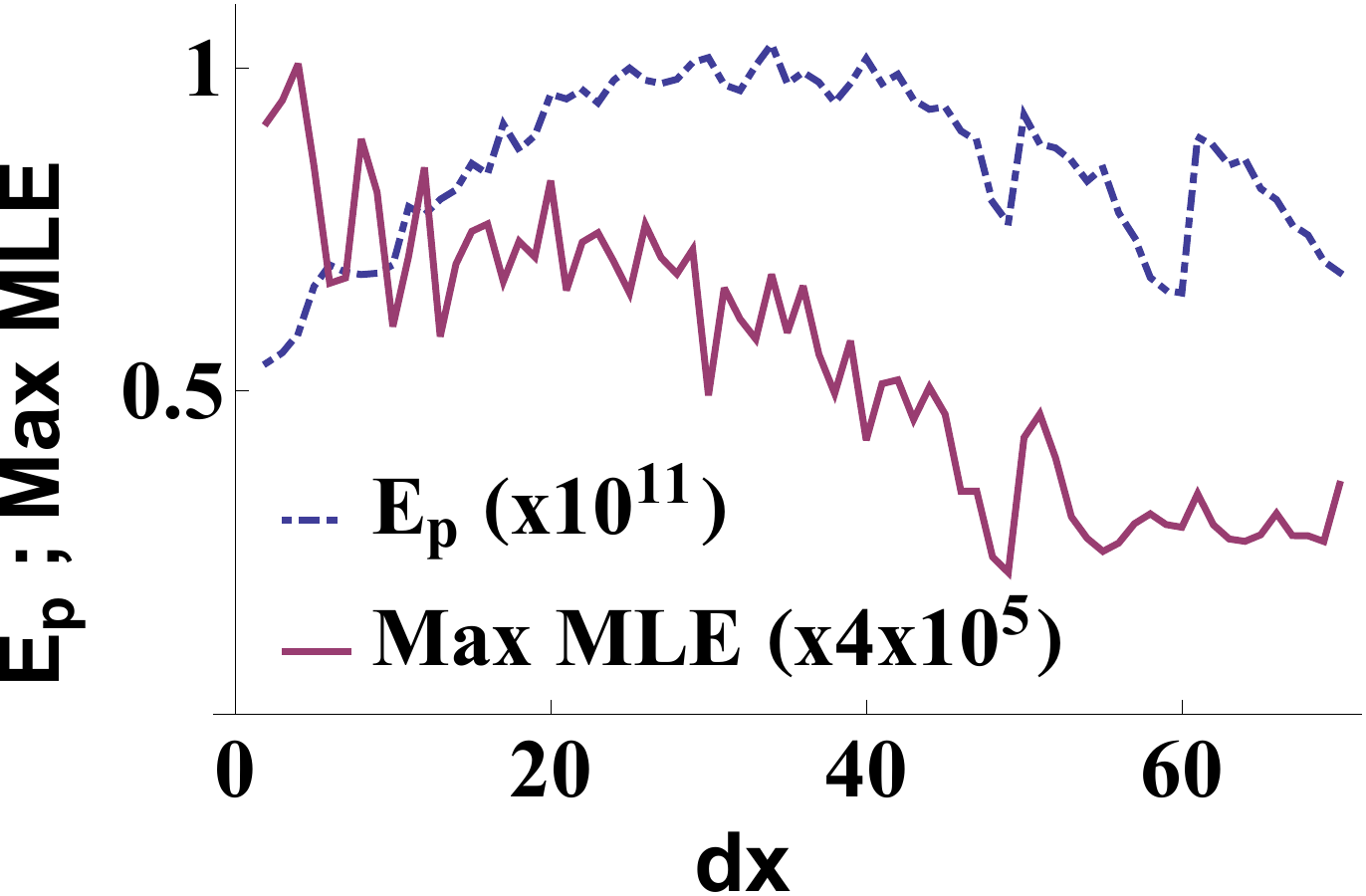}
\caption[PE and Max MLE versus fuelling rate - running model]{$E_p$ (left axis) and Max MLE (right axis) versus $dx$ for the running model. Max MLE size decreases with increasing $dx$, while $E_p$ peaks at about $dx=35$ (i.e. $dx/Z_c=0.3$). Both $E_p$ and Max MLE size show elements of fine structure, although changes in $E_p$ are more pronounced.}
\label{fig:pe-and-inverse-max-mle-versus-fuelling-rate}
\end{figure}

Figure \ref{fig:pe-and-inverse-max-mle-versus-fuelling-rate} shows the dependency between driving ($dx$) and maximum MLE size. Over the subinterval $0<dx<35$ there is a systematic increase of $E_p$ with increasing $dx$, while maximum MLE size falls in the same subinterval. For $dx>35$, both $E_p$ and maximum MLE size generally fall, while step changes in maximum MLE size coincide with changes in $E_p$, although the direction of correlation is not constant. 

\begin{figure}
\centering
\includegraphics[width=\linewidth]{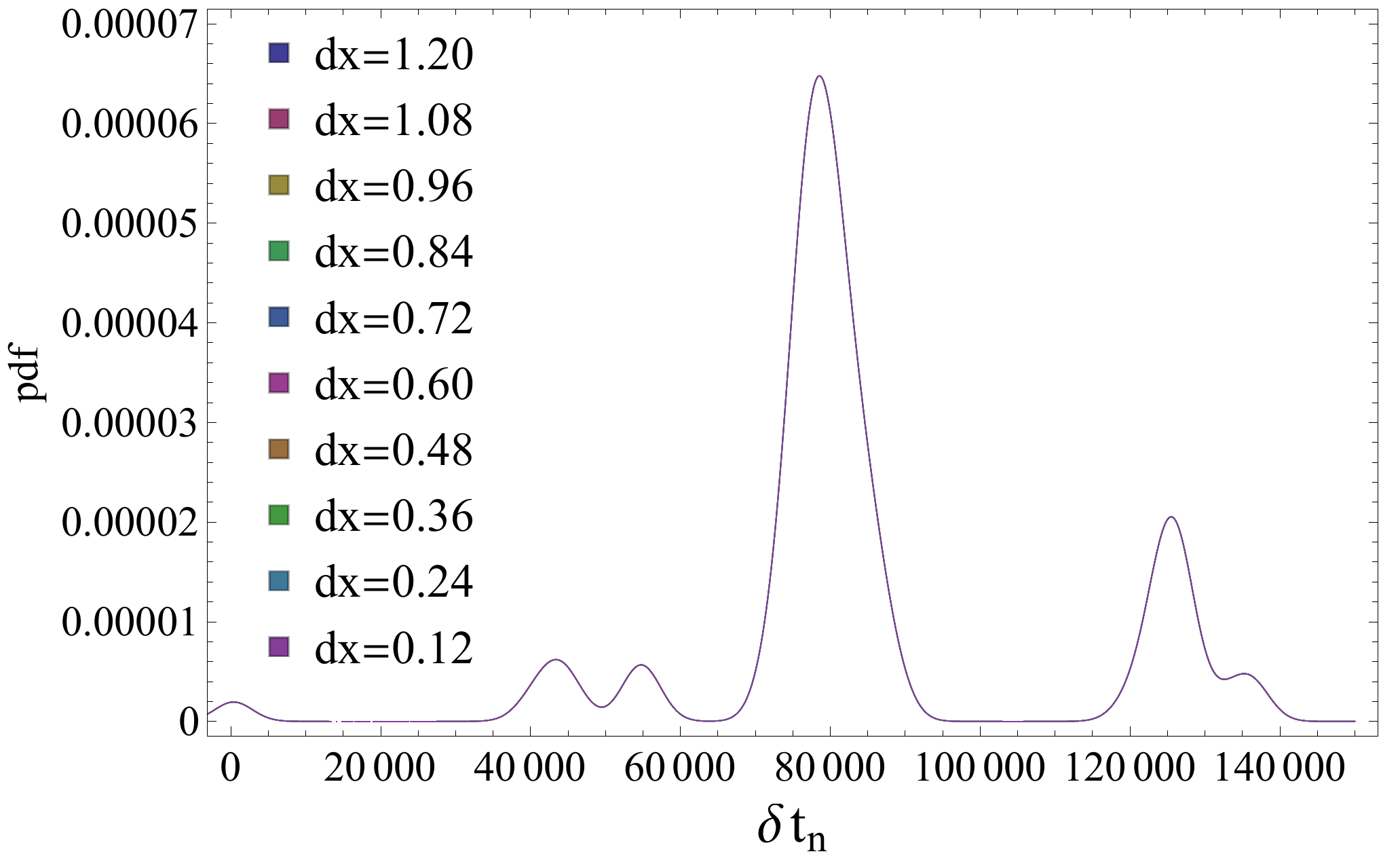}
\caption[Classic avalanche - $dx/zc = 0.01$]{Combined pdfs of waiting times between MLEs (classic model) $dx/Z_c = 0.01$ for $dx=0.12$ to $1.20$ in increments of $0.12$, and $Z_c=12$ to $120$ in increments of $12$. The PDFs overlap entirely, suggesting that waiting times are identical for identical values of $dx/Z_c$, regardless of the specific values of $dx$ or $Z_c$.}
\label{fig:Instantavalanche-dx_zc_point01}
\end{figure}

We have observed that the pdf of waiting times between MLEs also overlaps for common values of $dx/Z_c$, keeping $L_f$ constant.
Figure \ref{fig:Instantavalanche-dx_zc_point01} shows combined waiting time pdfs in the classic model for a fixed value of $dx/Z_c$, by incrementing both $dx$ (upwards) and $Z_c$ (downwards). Although ten pdfs are shown, they overlap to the extent that only one line is observed. Figure~\ref{fig:Instantavalanche-dx_zc_point01} shows that, for this model, $dx/Z_c$ influences waiting time behaviour, while the specific values of $dx$ and $Z_c$ are not relevant. As a result, it is necessary only to vary either $dx$ or $Z_c$, and not both. Here we consider variations only of $dx$.

We also observe that while all significant changes appear to correspond with integer ratios, not all integer ratios correspond with significant changes. To the best of the authors' knowledge, this integer ratio behaviour at high driving is a new result which has not previously been reported.

\begin{figure}
\centering
\includegraphics[trim= 0mm 0mm 0mm 0mm,clip,width=\linewidth]{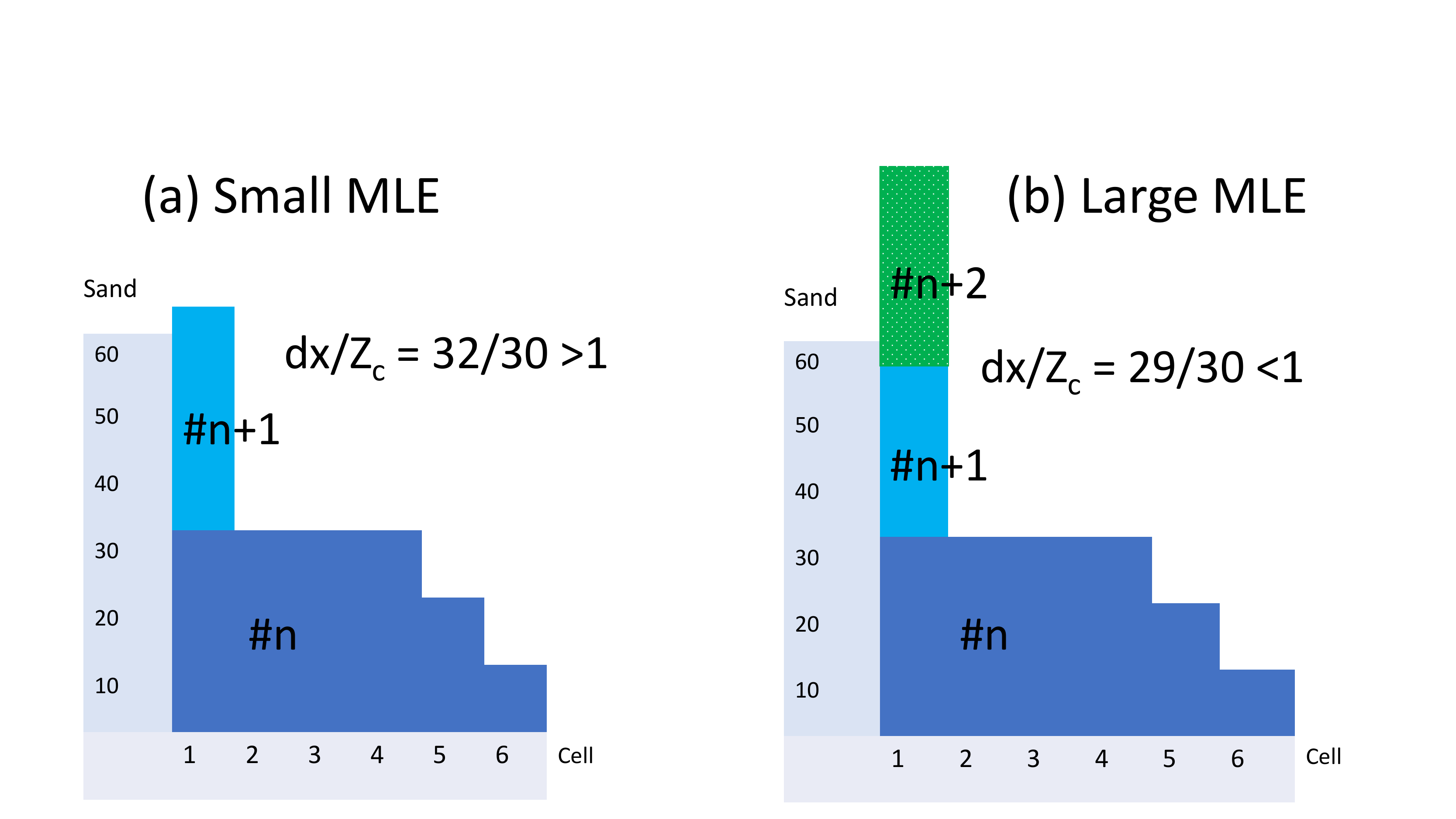}
\caption[Avalanche explanation]{Schematic sandpile at iteration prior to MLE for (a) $dx=32, Z_{c} =30$. Iteration no. $n+1$ will trigger an avalanche, as one iteration is enough for gradient to exceed $Z_{c}$. 32 grains to be distributed.(b) 
$dx=29, Z_{c}=30$. Iteration no. $n+2$ will trigger an avalanche, as two iterations required for gradient to exceed $Z_{c}$. 58 grains to be distributed.}
\label{fig:Addedsand-criticalgradient}
\end{figure}

A qualitative explanation for this behaviour may be suggested as follows. As shown in Figure~\ref{fig:Addedsand-criticalgradient}, the amount of sand to be distributed at each time step will increase as $dx$ increases, but will decrease just at or after integer ratios of $dx/Z_{c}$. In a particular example, the point at which the amount of sand to be distributed increases or decreases will also be dependent upon the local gradient of the sandpile at that time, i.e. it is the difference between the critical gradient and the gradient at a particular location in the sandpile which is relevant. In the example, the sandpile in Figure \ref{fig:Addedsand-criticalgradient}(a) at iteration no. $n+1$ will experience an avalanche, as $dx>Z_c$; whereas the sandpile in Figure \ref{fig:Addedsand-criticalgradient}(b) will undergo an avalanche at iteration no. $n+2$. % Although the sandpile may have a non-zero average gradient, it will nonetheless commonly be the case that adjoining cells take on identical values, as a result of avalanches which have not caused the critical gradient to be exceeded at all cells, even though they may have propagated entirely through the sandpile. From the perspective of such adjoining cells, the total amount of sand added to the system may not result in the critical gradient being exceeded at those points, even though the sum of the average gradient and $dx$ is greater than the critical gradient at other points in the sandpile. 
 The closer the actual gradient to the critical gradient following an avalanche, the lesser the amount of sand which must be added to trigger the following avalanche. The larger the amount of sand to be redistributed in a particular avalanche, the less likely it is that the sand will be assimilated within the sandpile, rather than causing a systemwide avalanche~\cite{Watkins1999}. 

This heuristic explanation suggests that the behaviour may be observable in real world scenarios involving large discrete fuelling.

\section{Limits of models at very high driving \label{very high driving - running}}

We now explore the limits of the models at very high driving, in both the running and classic models. We observe that in the running model, the algorithm becomes exactly solvable at very high driving. We discuss below the solution to the algorithm at very high driving, and the conditions under which this solution is valid.

\begin{figure}
\centering
\includegraphics[width=1\linewidth,trim={0cm, 0cm, 0cm, 0cm}, clip]{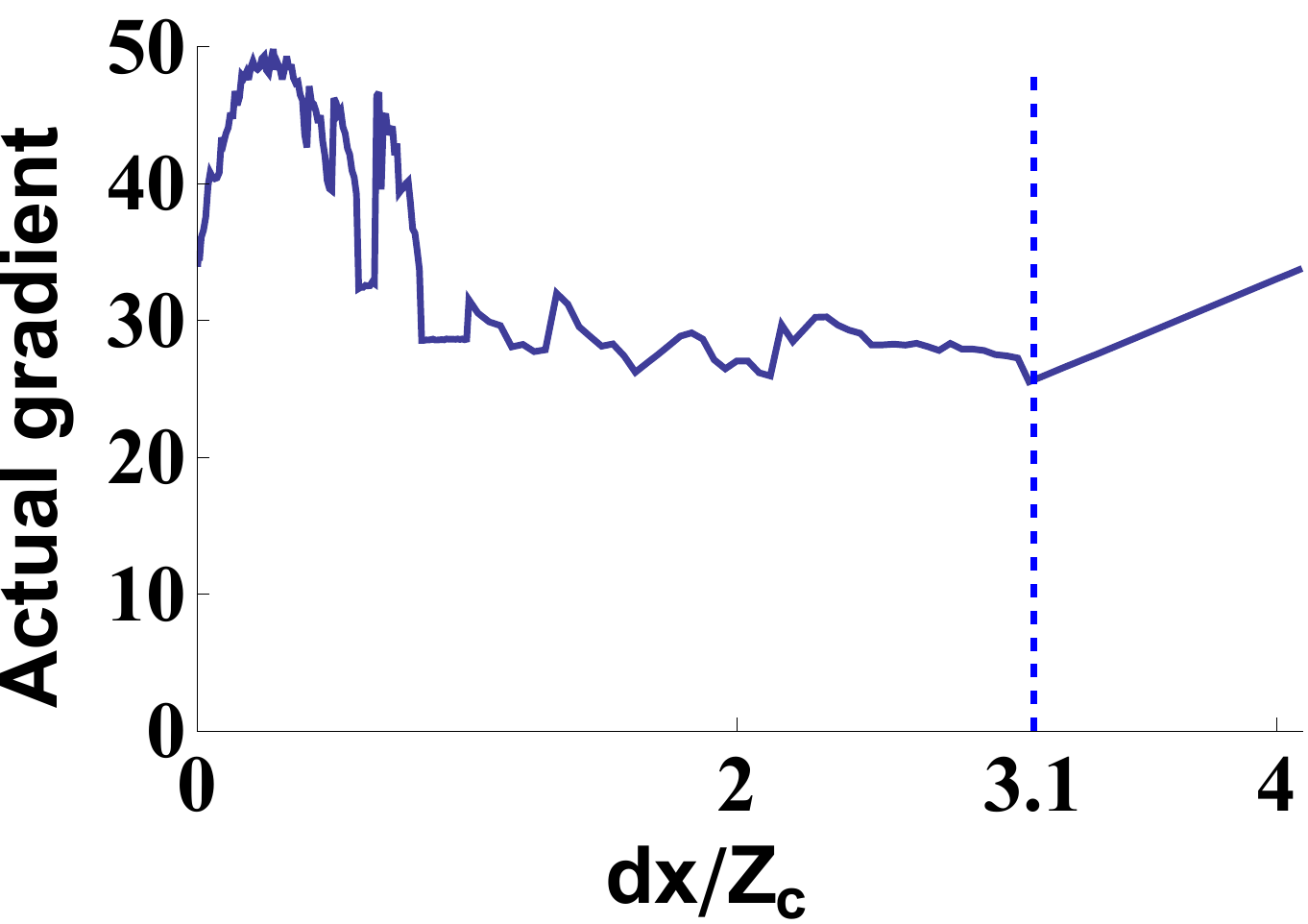}
\caption[PE driving up to 490]{Actual gradient (Z) of the resulting sandpile in steady state, as a function of $dx/Z_c$, for driving up to $dx/Z_c\approx 4.1$. Elements of fine structure are observed up to $dx/Z_c\approx 3.1$ after which $Z$ increases linearly with $dx/Z_c$. The actual gradient of the sandpile is closely related to $E_p$, as the total size of the sandpile is determined by its gradient. It is also related to $E_p/E_{p Max}$. We have shown actual gradient, rather than $E_p/E_{p Max}$ in order to show the straight line relationship between $dx/Z_c$ and actual gradient for values of $dx/Z_c>3.1$. }
\label{fig:System size-veryhighdriving-upto490}
\end{figure}

As shown in Figure \ref{fig:System size-veryhighdriving-upto490}, further increases in driving in the running model lead to an inflection point beyond which the relationship between driving and system size is approximately linear. This linear relationship continues indefinitely, regardless of the extent to which driving is increased.

\begin{figure}
\centering
\includegraphics[width=1\linewidth,trim={0cm, 0cm, 0cm, 0cm}, clip]{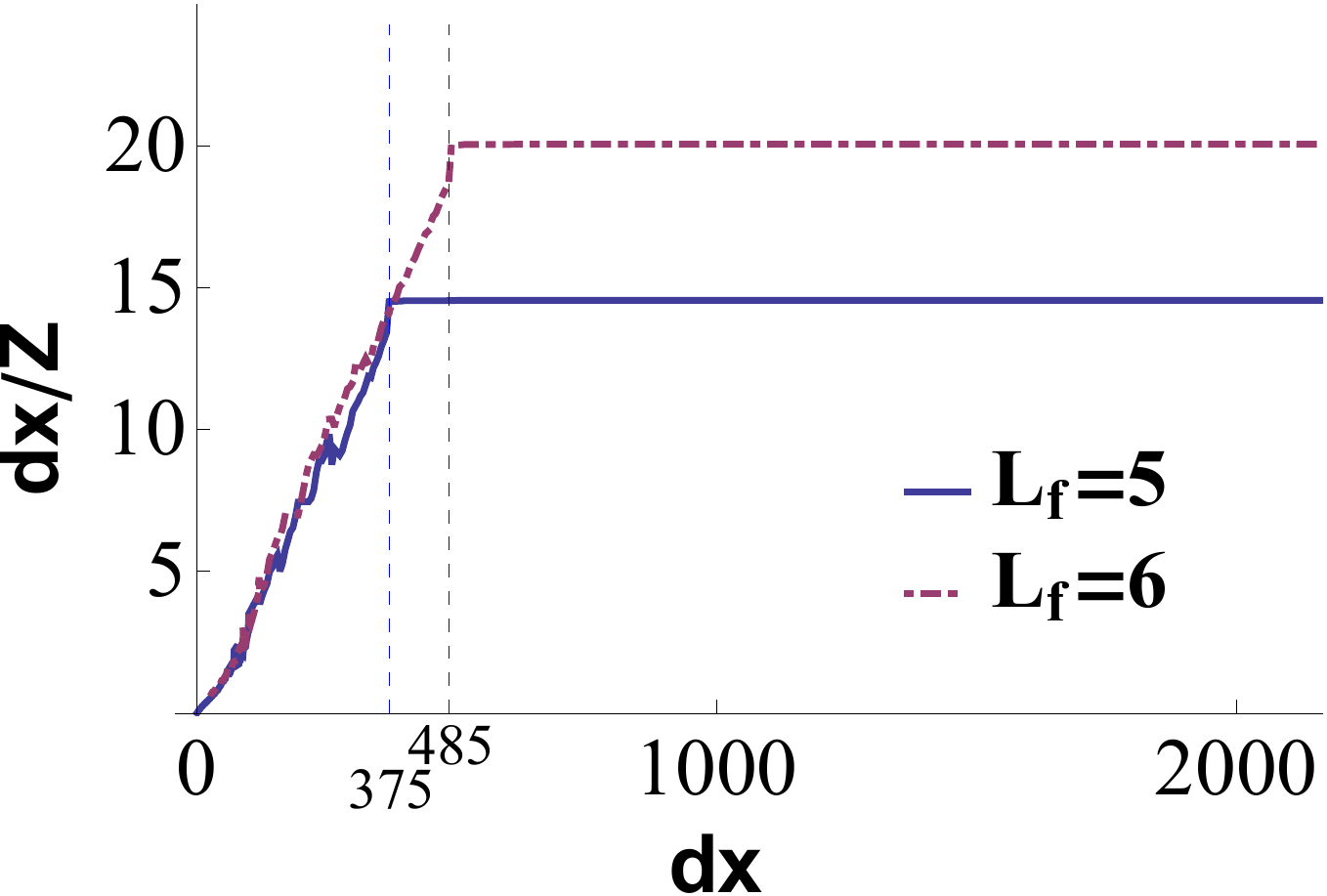}
\caption[PE driving up to 490]{$dx/Z$ as a function of $dx$: $L_f=5$ and $L_f=6$}
\label{fig:limitations on gradient-multiple Lf}
\end{figure}

Figure \ref{fig:limitations on gradient-multiple Lf} shows that the linear relationship, for $L_{f}=5$, occurs for values of $dx/Z$ (where $Z$ is the actual gradient) slightly less than $15$, and for $L_f=6$, for values slightly below $21$. These relationships are explained below, and indeed, for very high driving (which will be discussed further below), the behaviour and values of the sandpile at steady state are completely predictable. The behavior is predicted precisely for a given value of $dx$ using a simple formula beginning at the edge (RHS), and can also be largely explained using a simple formula beginning at the core (LHS).

%from Potential energy versus rate of added sand - Combined - Jan 2019. Tab 'Lf changes' Sixth chart from left

At very high driving, the system is completely stable, as the amount of sand injected at each timestep is exactly equal to the amount of sand lost at each timestep, and further, the value of the each cell can be determined analytically and remains stable at that analytic value.

The system will be in a stable state if the last $L_{f}$ cells at the edge (RHS) are equal in value to each other and to $dx$. If this occurs, each of those cells will also be equal in value to the amount of sand lost at each timestep, as the fluidization formula will equally distribute sand across the last $L_{f}$ cells plus the following cell, which is the cell at which the sand is lost. Stability will be achieved because the amount of sand entering the system at each timestep equals the amount of sand leaving the system at each timestep. The sandpile is exactly solvable in this state as $x_n$, the amount of sand in cell $n$, is in each case able to be determined using the values of cells $x_{n+1}$ to $x_{n+L_f+1}$. This is because at each cell the avalanche evolves by distributing sand across $L_{f}+1$ cells, meaning that at each following step $1/L_{f}$ of the difference between the cells has been left behind. Each of these $1/L_{f}$ is totalled to give the difference between cells.  As a result, the value of each cell (${x_n}$), where $n<(N-L_f)$, is given by:
\begin{equation}
x_{n}=x_{n+1}+\dfrac{x_{n+1}-x_{n+L_f+1}}{L_f}
\label{RHSvalue}
\end{equation}
For $n\geq(N-L_f)$, $x_n=dx$ as the last $L_f$ cells will all be equal to each other and to $dx$.

We can also determine the ratio between the actual gradient, $Z$, and $dx$ by first considering the core (LHS) of the sandpile. If there is a value of $dx$ for which the system enters a steady state (i.e. the number of particles added equals the number of particles lost at each time step, and the shape of the sandpile also remains unchanged), then each cell will have the same value before and after the addition of sand to the system. The amount of sand in cell $1$ (at the core or LHS of the system) will change when particles are added, and as the avalanche propagates through the system. If $Z$ is constant, the first $L_{f}+1$ cells will have the following values prior to the addition of sand: {$NZ,(N-1)Z,... ,(N-L_{f})Z$}. Following the addition of sand, and assuming that the amount of sand introduced is sufficient to trigger an avalanche which runs for at least $L_f$ steps, then after $L_{f}$ steps of the avalanche the values of the $L_{f}+1$ cells will be equalised, so that each cell will take on the average of the initial values of those cells, plus $dx/(L_{f}+1)$. At this point in the avalanche, the value of each of the first $L_f+1$ cells will be identical and given by:

\[
\dfrac{(nZ + (n-1)Z+...+ (n -L_{f})Z)}{(L_{f}+1)} + \dfrac {dx}{(L_{f}+1)}
\]

which, after gathering the $nZ$ terms, reduces to:
\[\dfrac{ nZ(L_{f}+1) - (1+2+...+L_{f}Z)}{(L_{f}+1)}  + \dfrac{dx}{(L_{f}+1)}\]

Cell $1$ will retain this value as the avalanche propagates to cell $L_f+1$, although the values of the following cells will change as the avalanche propagates (and indeed, will revert to their former values prior to the addition of sand). The amount of sand in cell $1$ will be unchanged before and after the addition of $dx$ if, prior to the addition of sand, $nZ$ is given by:

\[
nZ = \dfrac{ nZ(L_{f}+1) - (1+2+...+L_{f}Z)}{(L_{f}+1)}  + \dfrac{dx}{(L_{f}+1)}
\]

By manipulation of this equation, we can then determine the necessary value of $dx$ in terms of $L_f$ and $Z$:

\[\dfrac{1 + L_f}{2}L_fZ=dx\]

For $L_{f} = 5,\dfrac{1 + L_f}{2}L_f=15$, therefore if $15 Z = dx$, the value of the first cell will remain unchanged. If this is true for cell $1$, it will also be true for all other cells $n$, other than for those cells at the RHS, which is discussed above. 

For $dx=4000$, $L_{f}=5$, the difference between cells $1$ and $2$, as determined by iterating the
 running sandpile model to stability using these values, is 266.66, i.e. $15/4000$, which confirms the above result. As observed above, this value is also given by the formula beginning at the edge (RHS). 

\begin{figure}
\centering
\includegraphics[width=1\linewidth]{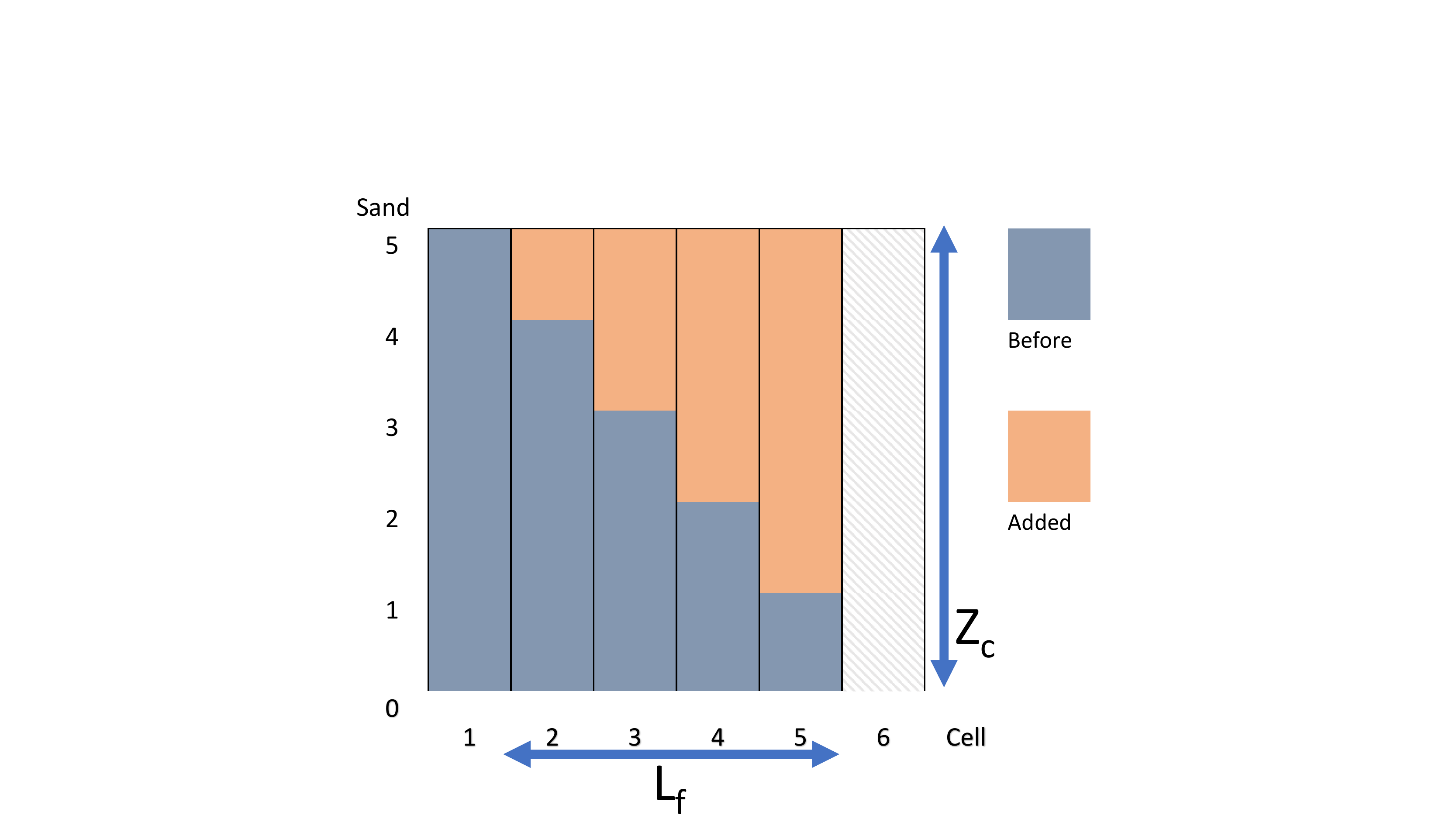}
\caption[$dx$ necessary to trigger systemwide avalanche in steady state]{$dx$ necessary to trigger systemwide avalanche in steady state \mody for $L_f=4$. \norm}
\label{fig:critical-gradient-and-added-sand-figure}
\end{figure}

The requirement that $dx/Z=15$ is a necessary, but not sufficient, condition, which is not dependent on $Z_c$. However, there is a further necessary condition mentioned above which is dependent upon $Z_c$, namely that the avalanche must \mody continue to \norm propagate \mody such that it reaches cell \norm $L_f+1$. For this to occur, the additional sand added at each time step must exceed the sum of the differences between $Z$ and $Z_c$ in each of the first $L_f+1$ cells. We show this schematically in Figure~\ref{fig:critical-gradient-and-added-sand-figure}. The area shaded in blue represents the state of the sandpile prior to the addition of $dx$, with the difference between each cell equal to $Z$. When $dx$ is added, an avalanche results which propagates through the first $L_f+1$ cells so that each the value of each cell is equal to the initial value of cell $n=1$ (which is a necessary consequence of the process of fluidizing). If the amount of sand which has been added ($dx$) is slightly greater than that necessary to reach the state in which the first $L_f+1$ cells are equal to the initial value of cell $n=1$, then the avalanche will continue to propagate and our condition will be met.

Figure~\ref{fig:critical-gradient-and-added-sand-figure} shows that the total amount to be added is equal to the total amount by which the cells from \mody $2$ \norm to \mody $5$ \norm exceed the height of cell \mody $6$ \norm - i.e. the amount which is necessary to complete the square shown in Figure~\ref{fig:critical-gradient-and-added-sand-figure}. The step height of the blue area is, by definition, equal to $Z$, the actual gradient. \mody The pink area \norm is given by the sum of cells \mody $2$ to $5$ \norm, multiplied by $Z$ which is $\dfrac{1 + L_f}{2}L_fZ$ (i.e. our formula above for $dx$). If $Z=1$, and \mody $L_f=4$, \norm this means that the amount of sand to be added is \mody $10Z$ \norm, as shown above. We can also see that cell $6$, which is equal in height to $Z_c$, \mody would contain, if filled, \norm $5Z$, which is \mody $dx/2$ \norm. This means that our second condition will be satisfied where $dx$ is >$2\times Z_c$. 

The condition can be generalised for all values of $L_f$. In order to complete the square in Figure~\ref{fig:critical-gradient-and-added-sand-figure},  $dx = \dfrac{1 + L_f}{2}L_fZ$, and the value of cell $L_f+1$ (which is equal to $Z_c$) is \mody $(L_f+1)Z$ \norm. Substituting we get \mody  $\dfrac{dx}{Z_c}=\dfrac{(1 + L_f)/2(L_fZ)}{(L_f+1)Z}$ \norm. 
Simplifying and rearranging gives us \mody $dx = L_fZ_c/2$. \norm The avalanche will propagate if $dx$ exceeds this value, so that our second condition is that \mody $dx > L_fZ_c/2$ \norm.

Our two conditions are then that $dx=\dfrac{1 + L_f}{2}L_fZ$ and \mody $dx >\dfrac{L_fZ_c}{2}$. \norm If both conditions are satisfied, then the system can continuously avalanche, and the total system size will be given by the sum of the cells calculated as above.

%from pdf and powerpoint files of the same name

Figure \ref{fig:limitations on gradient-multiple Lf} shows that the inflection point for $Z_c =120, L_f=5$ occurs at $dx=370$, which is greater than \mody $300$\norm, and that for $Z_c=120, L_f=6$, the inflection point occurs at $dx=490$, which is greater than \mody $360$\norm. \mody For $Z_c =120, L_f=1$ (not shown), the inflection point occurs at $dx=61$, which is greater than $60$. \norm

As with our observations in relation to the shape of the $E_p/E_{p Max}$ curve at medium values of $dx$ (Figure \ref{fig:PE_MaxPE}) and the pdf of MLE waiting times (Figure \ref{fig:Instantavalanche-dx_zc_point01}), the key is the relationship between $dx$ and $Z_c$, not their absolute values.

\begin{figure}
\centering
\includegraphics[width=1\linewidth,trim={0cm, 0cm, 0cm, 0cm}, clip]{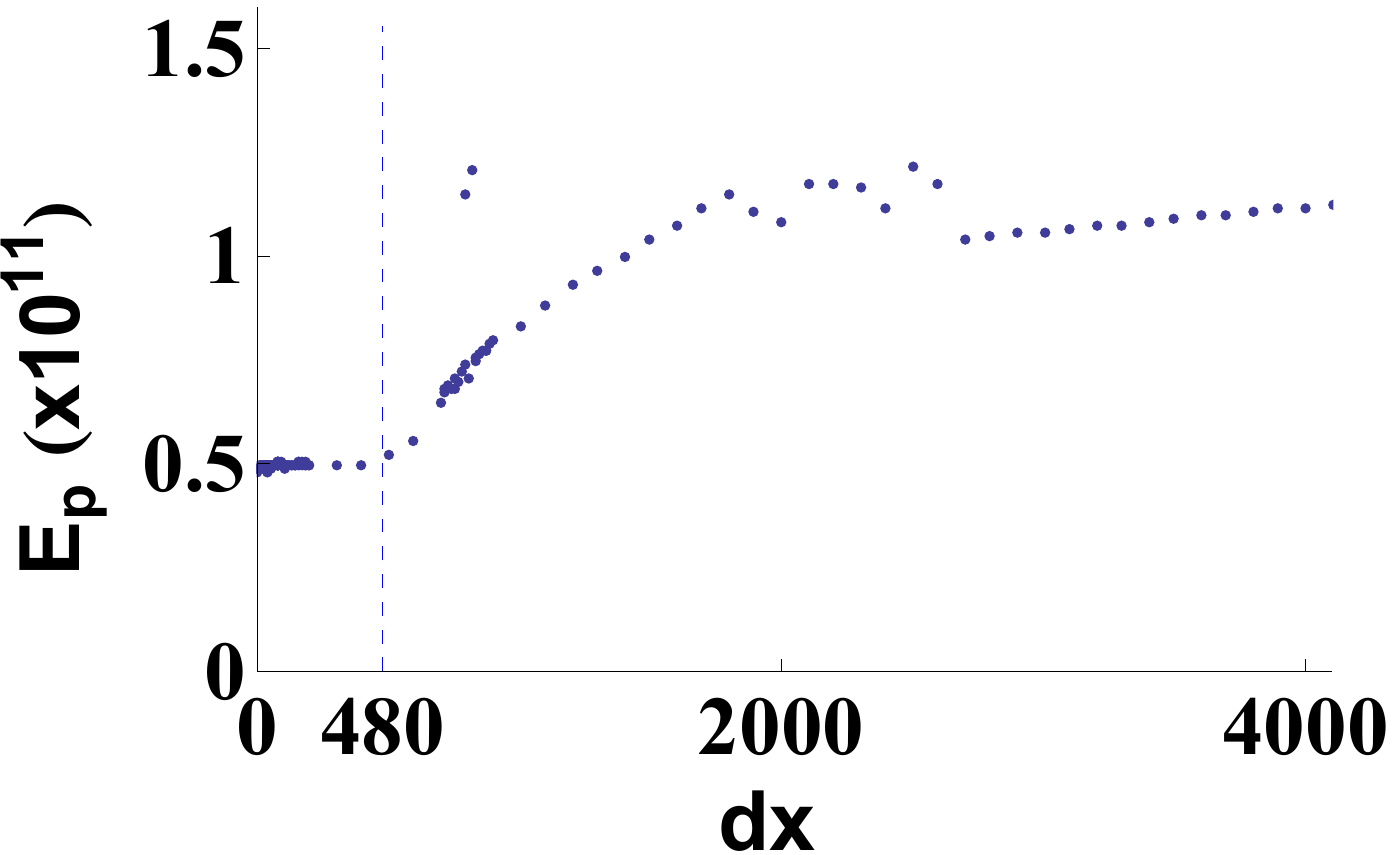}
\caption[$E_p$ - up to $dx = 4100$]{Classic model - $E_p$ as a function of $dx$, up to $dx = 4100$, for $L_f=5$. $E_p$ remains constant up to $dx\approx 480$, after which elements of fine structure appear.}
\label{fig:Instantavalanche-PE-uptodx=4100}
\end{figure}

We have also considered the behaviour of the classic model at very high driving. Typically, parameters for the classic model are set such that $dx/Z_c\approx0.01$. The $E_p$ of the classic sandpile remains constant as driving increases up to $dx/Z_c\approx3.3$, a range of two orders of magnitude. Above this value, non-linear behaviour is observed, as shown in Figure \ref{fig:Instantavalanche-PE-uptodx=4100}.

%from Relationship between fuelling and waiting time peaks-copy. Tab 'Potential Energy - instant'

The cause of the non-linear behaviour at very high driving can be attributed to the fact that the sandpile is swept only once between each addition of sand, in the absence of a systemwide avalanche. If $dx$ is high enough, sand is not swept from the core to the edge before further sand is added. When driving becomes high enough, the gradient at the core (LHS) exceeds $Z_c$. 

We also observe that there are a couple of anomalous data points, at $dx=800,dx=820$, where the sandpile demonstrates the high driving behaviour of the running model, namely that the amount of sand exiting the sandpile at each timestep is equal to $dx$. The actual gradient ($Z$) for these anomalous cases is also consistent with the results for the running model. For example, at $dx=820$, $Z=54.667$ so that $dx/Z=15$, which is the same result as obtained for the running model. 

Increasing the driving rate for the classic model to the extent that sand is not fully swept across the system between iterations contradicts the central premise of the classic model, which operates on the assumption that the sandpile fully relaxes before further sand is added. We have therefore not discussed in detail the behaviour of the classic model at very high driving. It is sufficient to observe that $E_p$ in the classic model is unaffected while the model continues to obey its central premise, that of full relaxation before sand is added. As discussed above, MLE size is also unaffected in this regime, while waiting times reduce as driving increases.

\vspace{-0.25cm}
\section{Conclusions \label{sec:conclusions}}

We have employed a simple sandpile model that emulates pellet pacing in a fusion plasma. \mody The model is a 1D centrally fuelled non-conservative sandpile which exhibits avalanching behaviour (noting that while 2D models might also be considered, they are not our focus here). \norm  While the model is typically used at low constant driving, we have adapted it in two alternative ways: firstly by providing for high constant driving, and secondly by adding intermittent `pellets' of sand at the core concurrent with low constant driving. We have employed two versions of the model, a classic model in which fuelling is paused during a systemwide avalanche, and a running model in which fuelling continues during an avalanche.

 In the low to medium constant driving regime, average potential energy in steady state ($E_p$) varies with $dx$ in the running model, while it remains constant with changes in $dx$ in the classic model. For the running model, analysis of $E_p/E_{p Max}$ against $dx/Z_c$ for increasing $dx$ shows that step changes occur, often at integer ratios. A heuristic explanation is suggested for this behaviour. At very high constant driving, the behaviour of the running model can be analysed, such that the exact value of each cell of the sandpile can be determined analytically given the value at cell $n=1$. For the classic model, $E_p$ increases with $dx$ at very high driving. This behaviour appears to arise as significant fuelling occurs during non-systemwide avalanches, which is inconsistent with the central premise of the classic model that the system should relax to stability between timesteps.
 
 In the intermittent fuelling regime, $E_p$ slowly increases with pellet size, while max MLE size increases more quickly. By contrast, in the constant fuelling regime, using the running model, $E_p/E_{p Max}$ increases up to about $dx/Z_c\approx0.3$, then slowly decreases, while max MLE size slowly decreases through this range.
 
This suggests that MLE control is more successful with increased constant fuelling, than with intermittent fuelling.
 
We can compare these results to pellet pacing in fusion plasmas. For example, in ELM pacing experiments at JET, while an increase in ELM frequency was observed, which might be expected to reduce ELM size, virtually no reduction in peak heat flux was observed \cite{Lang2013B, Lang2015}. This lack of reduction in peak heat flux appears consistent with our results which suggest that max MLE size is not reduced by the introduction of pellets in our sandpile model.

An aspect of pellet pacing in fusion plasmas which is not captured in our model is that pellets for ELM mitigation are typically added at the top of the pedestal - pellets injected into the core are typically used for fuelling rather than pellet pacing. In future work, we propose to adapt a version of the sandpile model which incorporates a pedestal \cite{Bowie2018} to determine whether that will produce a better comparison with experimental results. Longer term, the extension to a 2D sandpile offers the possibility of capturing the radial-varying poloidal and toroidal twist of the magnetic field. Such a model could capture this dependence by making the second dimension a periodic poloidal angle, in which the poloidal angle increment is non-uniformly spaced in radius. The model would thus capture sandpile transport for spatially-localised sand-grains that take the magnetic topology to the plasma edge.  Such transport might be a proxy for radially localised modes. It is unclear how such a 2D model might capture transport from modes with global radial extent.

%\vspace{-0.5cm}
\section*{Acknowledgments}
%\vspace{-0.5cm}
%\noindent 
The authors would like to thank an anonymous referee for their very helpful suggestions. This work was jointly funded by the Australian Research Council through grant FT0991899 and the Australian National University. One of the authors, C. A. Bowie, is supported through an ANU PhD scholarship, an Australian Government Research Training Program (RTP) Scholarship, and an Australian Institute of Nuclear Science and Engineering Postgraduate Research Award.

\vspace{-1.5cm}
\renewcommand\refname{}
%the command above removes the section heading references
%\bibliography{highdriving}

\end{document}